\documentclass[a4paper,11pt]{article}
\usepackage[dvips]{graphicx}
\usepackage{amssymb}
\usepackage{amsmath}

\usepackage{cite}

\begin{document}

\title{The 2d Gross-Neveu Model for Pseudovector Fermions and Tachyonic Mass Generation}
\author{V. K. Oikonomou\thanks{
voiko@physics.auth.gr}\\
Department of Theoretical Physics, Aristotle University of Thessaloniki,\\
Thessaloniki 541 24 Greece} \maketitle

\bigskip

\bigskip

\bigskip

\begin{abstract}
Recent observations in the OPERA experiment suggest that the
neutrino could propagate with speed that is superluminal. Based on
early theoretical work on tachyonic fermions we shall study a
modification of the Gross-Neveu model in two dimensions. We shall
see that the theory results to the dynamical generation of real
and imaginary masses. These imaginary masses indicate the
possibility that tachyonic solutions (or instabilities) could
exist in the theory. The implications of a tachyonic neutrino
coming from astrophysical sources are critically discussed.
Moreover, we present a toy model that consists of an $U(2,2)$
invariant Dirac Lagrangian. This theory can have tachyonic masses
as solutions. A natural mass splitting between the solutions is a
natural outcome of the formalism.
\end{abstract}

\pagebreak

\section*{Modified 2d Gross-Neveu Model and Extrema of the Effective Potential}

The Gross-Neveu model describes Dirac fermions in 1+1 dimensions,
that have four fermions interactions, thus being a very elegant
and useful tool for studies of the strong interactions. It is
based on non-perturbative techniques, namely the large N expansion
and it results to the dynamical generation of a fermionic mass, a
phenomenon known as dimensional transmutation. It is described by
the Lagrangian density
\begin{equation}\label{gross}
\mathcal{L}=\sum_i^Ni\bar{\psi}_{i}\gamma^{\mu}\partial_{\mu}\psi_i+\frac{1}{2}g_0^2\Big{(}\sum_i^N\bar{\psi}_i\psi_i\Big{)}^2
\end{equation}
for $i=1,2,...N$. The Lagrangian (\ref{gross}) is invariant under
the chiral symmetry $\psi_i\rightarrow \gamma_5\psi_i$. This
symmetry is a kind of $Z_2$ symmetry, since $\gamma_5^2=1$. Due to
one-loop quantum corrections, this symmetry is broken by the
theory.

Recently an experimental result has put into question the nature
of the neutrino. Particularly, the OPERA experiment \cite{opera}
resulted that the neutrino propagates in space with speed that
exceeds the speed of light. It is obvious that the experimental
results must be re-examined in order to be sure that this result
is valid. Nevertheless it is an intriguing result that deserves
some attention and if proven true, much of the theoretical work
involving the neutrino must be put in a new conceptual basis.
Actually it has draw the attention of many theorists
\cite{newentries,newentries1,newentries2,newentries3,newentries4,newentries5,newentries5a,newentries6,newentries7,newentries7a,newentries8,nicolaidis,newentries9,newentries10,newentries11}.

Insightful studies on the tachyonic nature of the neutrino was
provided in the 80's by A. Chodos et. al
\cite{chodos,chodos1,chodos2}. The authors proposed experiments
that could prove the tachyonic nature of the neutrino field. They
actually proposed the time of flight experiment, which is an
OPERA-like experiment. Up to now, several papers followed,
studying the tachyonic neutrino
\cite{tachyonneutrino,tachyonneutrino1,tachyonneutrino2,tachyonneutrino3,tachyonneutrino4,tachyonneutrino5,tachyonneutrino6,tachyonneutrino7,chinese,malakas}.
The Lagrangian that describes the tachyonic neutrino is given by
\cite{chodos}:
\begin{equation}\label{tachyonlagra}
\mathcal{L'}=i\bar{\psi}\gamma_5\gamma^{\mu}\partial_{\mu}\psi-m\bar{\psi}\psi
\end{equation}
In the massless case, at the Standard Model level, the terms
$i\bar{\psi}\gamma_5\gamma^{\mu}\partial_{\mu}\psi$ and
$i\bar{\psi}\gamma^{\mu}\partial_{\mu}\psi$, are
indistinguishable.

In this brief letter we shall modify the Gross-Neveu Lagrangian
according to the massless limit of the tachyonic Lagrangian
(\ref{tachyonlagra}). The modified Gross-Neveu Lagrangian is:
\begin{equation}\label{mgross}
\mathcal{L}=\sum_i^Ni\bar{\psi}_{i}\gamma_5\gamma^{\mu}\partial_{\mu}\psi_i+\frac{1}{2}g_0^2\Big{(}\sum_i^N\bar{\psi}_i\psi_i\Big{)}^2
\end{equation}
We shall dwell on the mass generation issues, and how the results
are modified when using the above Lagrangian. We must note that
the massless field theory of the tachyons does not suffer from the
conceptual problems that the massive case has. Indeed no
consistent Quantum field theory of tachyons exists up to now
\cite{chodos}, however it could be a good starting point to work
towards the consistent incorporation of tachyons in the successful
field theories. The Lagrangian (\ref{mgross}) is invariant under
the chiral transformation $\psi_i\rightarrow \gamma_5\psi_i$. We
now compute the effective potential corresponding to
(\ref{mgross}). Using very well known techniques \cite{largen} we
have:
\begin{equation}\label{effe}
\int\mathcal{D}\psi\mathcal{D}\bar{\psi}e^{-i\int\mathrm{d}x^DV_{eff}(\sigma)}=\int\mathcal{D}\psi\mathcal{D}\bar{\psi}\exp\Big{[}\int\mathrm{d}x^D\Big{(}i\bar{\psi}_{i}\gamma_5\gamma^{\mu}\partial_{\mu}\psi_i-\frac{1}{2}g_0^2\sigma^2-\mu\sigma\bar{\psi}_i\psi_i\Big{)}\Big{]}
\end{equation}
where $\sigma$ is a composite field in terms of which we shall
express the final expression of the effective potential. In the
above, $\mu$ is a free parameter to be used for dimensional
reasons and $D=2-2\epsilon$. After performing the Gaussian
integration over the fermions and taking the $\epsilon \rightarrow
0$ we obtain:
\begin{equation}\label{effe1}
e^{-i\int\mathrm{d}x^DV_{eff}(\sigma)}=\exp\Big{[}i\int
\mathrm{d}x^D
\Big{(}-\frac{1}{2g^2}\sigma^2+\frac{N}{4\pi}(\frac{1}{\epsilon}-\gamma+\ln
(4\pi))\sigma^4-\frac{N}{4\pi}\sigma^4\ln\frac{\sigma^4}{\mu^4}\Big{)}\Big{]}
\end{equation}
Using the $\mathrm{\overline{MS}}$ renormalization scheme to
subtract the poles, the effective potential reads:
\begin{equation}\label{efffectivepotential}
V_{eff}(\sigma)=\frac{1}{2g^2}\sigma^2+\frac{N}{4\pi}\sigma^4\Big{(}\ln\frac{\sigma^4}{\mu^4}-1\Big{)}
\end{equation}
Minimizing the effective potential in respect to the parameter
$\sigma$, we obtain the equation:
\begin{equation}\label{finalequation}
\frac{1}{2g^2}+\frac{N}{2\pi}\sigma^2\ln(\frac{\sigma^4}{\mu^4})=0
\end{equation}
The above equation can have four real roots of the form
$(x_1,-x_1)$ and ($x_2,-x_2$), with $x_1,x_2$ real positive
number. This result is similar to the usual Gross-Neveu result.
This means that $\sigma$ can take four real values, $\sigma=\pm
x_1$ and $\sigma=\pm x_2$. The field $\sigma$ is the minimum of
the effective potential, thus the quantum theory can have two
equivalent ground states.

\begin{figure}[ht!]
\begin{minipage}{18pc}
\includegraphics[width=20pc]{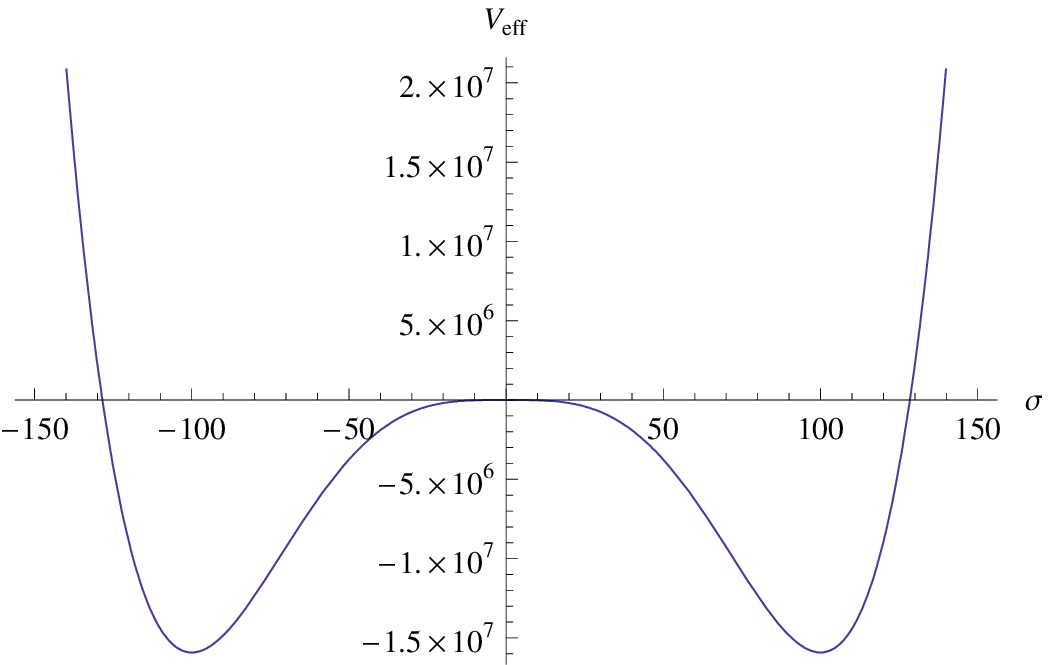}
\end{minipage}
\begin{minipage}{18pc}\hspace{2pc}%
\includegraphics[width=20pc]{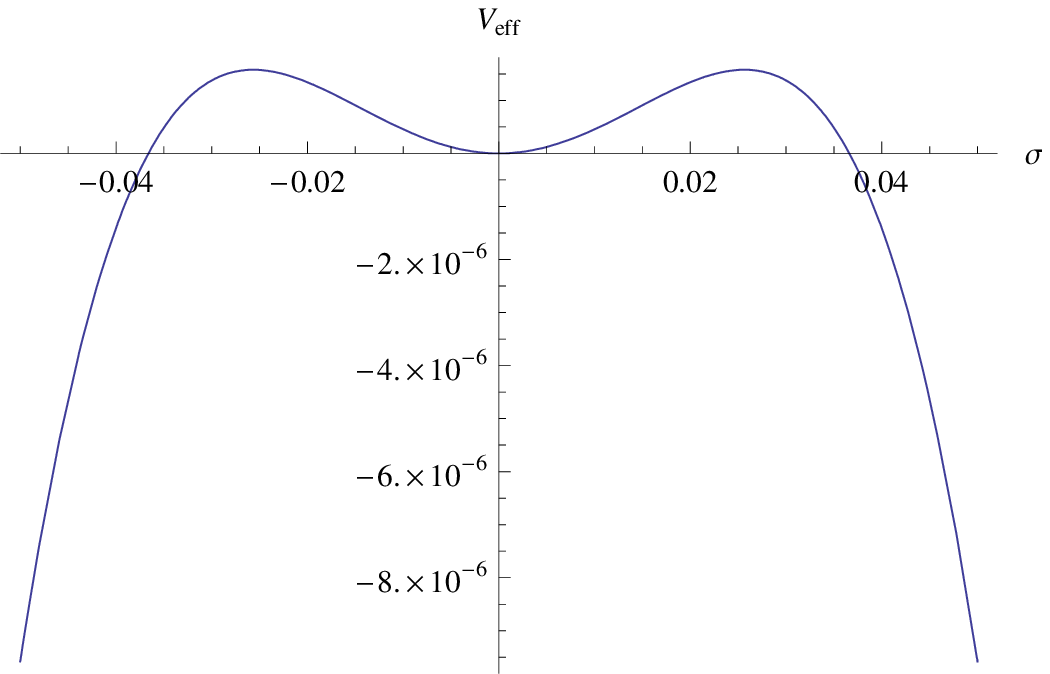}
\end{minipage}
\caption{The effective potential for $g^2N/\mu \sim 10$. In the
left figure we can see the minima of the potential at
$\sigma_1=100$ and $\sigma_2=-100$. In the right figure we can see
the maxima of the potential $\sigma_3=0.0091$, $\sigma_4=-0.0091$}
\label{neumtor}
\end{figure}

The fermions acquire finite masses which can be $m=|\sigma|=x_1$
and $m=|\sigma|=x_2$. It is obvious that the initial $Z_2$ chiral
symmetry is broken. Additionally, the fermionic fields can have
two different values for their mass, which is dynamically
generated. The modified Gross-Neveu model we studied results to
two equal masses (minima of the potential), a result similar to
the original Gross-Neveu result. But the most intriguing
difference in comparison to the Gross-Neveu is the existence of
complex extremal points of the effective potential. Indeed a
numerical study of the above equation (\ref{finalequation}) shows
that there are values of the expression $\frac{\pi\mu}{g^2N}$ for
which the equation has two complex roots. These roots indicate
instability of the quantum theory, since the mass-square of the
field $\sigma$ can be negative. But in view of the new
experimental data we could argue that the above model
(\ref{mgross}) can generate dynamically tachyonic masses.
Recalling that the modified version of the Gross-Neveu we studied
is based on a tachyonic Lagrangian, the fact that solutions for
tachyonic masses exist, can be very useful.

Let us give here some quantitative details of the calculation we
performed. Recall that the original two dimensional Gross-Neveu
model was described by the effective potential $V'_{eff}$,
\begin{equation}\label{originalgross}
V'_{eff}=\frac{1}{2g^2}\sigma^2+\frac{N}{4\pi}\sigma^2\Big{(}\ln\frac{\sigma^2}{\mu^2}-1\Big{)}
\end{equation}
The minimization process results to the following equation:
\begin{equation}\label{origrossnev}
\frac{1}{2g^2}+\frac{N}{4\pi}\ln \frac{\sigma^2}{\mu^2}=0
\end{equation}
It is obvious that equation (\ref{origrossnev}) has only two real
roots. In addition, if one tries to search for roots of the form
$\sigma=ix$, this results to mathematical inconsistencies. Hence
no imaginary roots exists for the original two dimensional
Gross-Neveu model. In the case of (\ref{finalequation}), we have a
plethora of roots. The most interesting are these of the form
$\sigma=ix$. Indeed, one can easily check numerically that if we
put $\sigma=ix$, then we wave two real solutions for $x$, namely
$x$ and $-x$. Hence the result is that,  equation
(\ref{finalequation}) has two purely imaginary roots, $\sigma=ix$
and $\sigma=-ix$ For example, if $Ng^2=10^{12}$ (large N-strong
coupling limit) and $\mu=10^6$, then $\sigma_1=10^{-7}i$ and
$\sigma_2=-10^{-7}i$. Therefore we have two equivalent tachyonic
mass vacuum state solutions. Apart from purely imaginary
solutions, equation (\ref{finalequation}) has complex roots of the
form $\sigma=a+bi$, and as we mentioned earlier real roots. Let us
discuss the implications of the complex roots at this point.
Complex masses are very closely related to tachyonic fermion
solutions. Indeed, a generalized form of Lagrangian
(\ref{mgross}), found in \cite{chinese}, results in the modified
Dirac equation \cite{chinese}:
\begin{equation}\label{mdiraceq}
(i\gamma^{\mu}\partial_{\mu}-\lambda_T)\psi=0
\end{equation}
with $\lambda_T=ia+b\gamma_5$. Under certain conditions, depending
on the $a$ and $b$ choices, one can obtain tachyonic solutions in
this case, but there are two more solutions. The authors of
\cite{chinese} call these ''luxons'' and ''bradyons''. To us,
these solutions are under serious investigation, since the
solutions lead to serious quantization problems, more severe than
the tachyonic solutions. In any case, if superluminal propagation
in spacetime is true, then we should work and find a consistent
quantization procedure, even by generalizing the theoretical
formalism of quantization. As for the real roots that equation
(\ref{finalequation}) has, the first set $\pm x_1$ are local
maxima and the other set $\pm x_2$ are local minima. Hence the
situation is equivalent to the 2d Gross-Neveu model, except the
two maximum solutions. In figures 1 (a) (left) and (b) (right) we
plot the effective potential (\ref{efffectivepotential}), for
$g^2N/\mu \sim 10$. We can clearly see the two minima,
$\sigma_1=100$, $\sigma_2=-100$ and the two maxima
$\sigma_3=0.0091$, $\sigma_4=-0.0091$. In figure 2 we have made a
plot of the roots corresponding to equation (\ref{finalequation}),
for various values of $g^2N/\mu$. The y-axis represents the
imaginary axis, while the x-axis, the real axis. As we can see
there are purely imaginary, complex and real roots.

\begin{figure}[h]
\begin{center}
\includegraphics[scale=1.3]{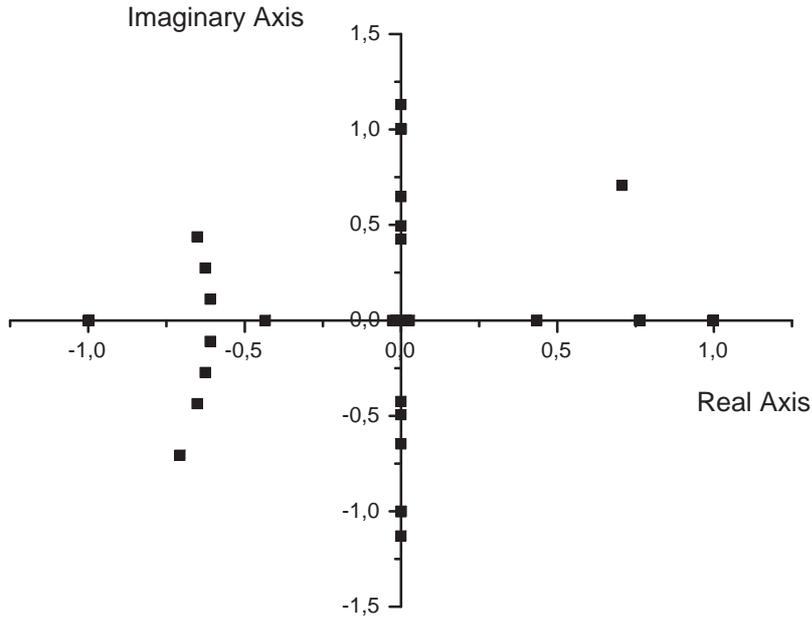}
\end{center}
\caption{The minima of the effective potential $V_{eff}$ for
various values of the parameters $g$, $\mu$ in the large $N$
limit} \label{roots}
\end{figure}

Before closing this investigation, let us study the effective
potential in various limiting cases of the parameters. We start
with the case $\frac{\sigma^4}{\mu^4}\ll 1$. Upon using the
approximation for the logarithm, valid for $x>0$:
\begin{equation}\label{approx}
\ln
x=2\sum_{k=1}^{\infty}\frac{1}{(2k-1)}\Big{(}\frac{x-1}{x+1}\Big{)}^{2k-1}
\end{equation}
and keeping the first term in the sum, the effective potential
$V_{eff}$ of equation (\ref{efffectivepotential}) can be cast as:
\begin{equation}\label{eff122}
V_{eff}(\sigma)=\frac{1}{2g^2}\sigma^2+\frac{N}{4\pi}\sigma^4\Big{(}\frac{\sigma^4}{\mu^4}-1\Big{)}
\end{equation}
If we minimize equation (\ref{eff122}) in respect to $\sigma^2$,
we obtain the following solutions:
\begin{align}
& \sigma_1=\frac{1}{30} \Big{(}m+\frac{g^2 N m^2}{A} +\frac{A}{g^2
N}\Big{)},
\\ \notag & \sigma_2=\frac{1}{60} \Big{(}2m+\frac{\big{(}-1-i
\sqrt{3}\big{)} g^2 N m^2}{A} +\frac{i \big{(}i+\sqrt{3}\big{)}
A}{g^2 N}\Big{)},
\\ \notag &
\sigma_3= \frac{1}{60} \Big{(}2 m+\frac{i \big{(}i+\sqrt{3}\big{)}
g^2 N m^2}{A} +\frac{\big{(}-1-i \sqrt{3}\big{)}A}{g^2 N}\Big{)}
\end{align}
where we introduced the parameter $A$, which is equal to:
\begin{equation}\label{alphaparameter}
A=\sqrt[3]{m^3 N^3 g^6-2700 m N^2 \pi g^4+30 \sqrt{6 \pi }
\sqrt{g^8 m^2 N^4 \Big{(}1350 \pi -g^2 m^2 N\Big{)}}}
\end{equation}
In the case we have $\frac{\sigma^4}{\mu^4}\gg 1$, we cannot have
analytic solutions. Nevertheless by using the approximation,
\begin{equation}\label{approxbig}
\ln x=-\mathrm{K}_0(x)-\gamma+\ln 2
\end{equation}
and in addition, the asymptotic form of the modified Bessel
function $\mathrm{K}_0(x)$, valid for large values of $x$:
\begin{equation}\label{approxbig1}
\mathrm{K}_0(x)\sim \sqrt{\frac{\pi}{2x}}e^{-x}
\end{equation}
we obtain a simpler expression for the effective potential. Indeed
the potential can be written as:
\begin{equation}\label{eff12223223}
V_{eff}(\sigma)=\frac{1}{2g^2}\sigma^2+\frac{N}{4\pi}\sigma^2\Big{(}\mu^2\sqrt{\frac{\pi}{2}}e^{-\frac{\sigma^4}{\mu^4}}-\gamma+\ln
2-2\Big{)}
\end{equation}

The original Gross-Neveu model was renormalization scale
invariant. We must say that the effective potential of the
modified Gross-Neveu is also $\mu$-independent, as can be easily
seen from the following equation
\begin{equation}\label{rge}
\mu^2\frac{\partial V_{eff}(\sigma)}{\partial \mu^2}=0
\end{equation}
We shall not pursuit further the renormalization group issues in
this paper.

\subsubsection*{A Discussion on Tachyonic Fermions and Their Impact on Physical Phenomena}

In this section we discuss some issues in reference to the
possible tachyonic nature of the neutrino. Obviously, the
experimental results must be further scrutinized in order to be
sure for their validity. However if true, the tachyonic nature of
the neutrino could alter some aspects of physical processes or
even give answers to unexplained mysteries. For example in
hot-dense backgrounds, the plasmon decay and the effective
neutrino-photon coupling must be reexamined in view of the
tachyonic nature of the neutrino. The modification of the
effective photon-neutrino interactions at high temperature could
modify the neutrino emission from stars, or the neutrino emission
from Gamma Ray Bursts. On that account, an interesting impact of
superluminal neutrino's is studied in \cite{newentries}. In
regards to Gamma Ray bursts, phenomenological results of quantum
gravity predict that neutrino's could arrive faster than the
photons \cite{alfaro}. This result is quite interesting since the
nature of the neutrino is controversial, and signals a strong
relation of this particle with gravity, in an undiscovered way. In
the 50's Wheeler \cite{wheeler} had noted that the neutrino could
be a geometrodynamical particle of some kind (in the next section
we shall present an Poincare invariance violating extension of the
Dirac equation, with interesting features nonetheless).

One outcome of the possible ``tachyonicity'' of the neutrinos is
that, in order to be consistent, ``tachyonicity'' requires the
existence of both right handed and left handed neutrinos. This is
a problem for the neutrino since we have observed only left handed
neutrinos. Nevertheless, this is already a problem of contemporary
neutrino physics, which is solved by the seesaw mechanism
\cite{seesaw}.

A negative mass square can be a very strange physical result.
However we must note that, the mass parameter even in classical
mechanics, is a parameter that cannot be measured directly even
for slow particles \cite{bilaniuk}. Only energy and momentum are
measured through interactions and therefore must be real
\cite{bilaniuk}. As noted by the authors of \cite{bilaniuk}, the
imaginary mass offends traditional thinking and not observable
physics. Nonetheless, imaginary mass particles cannot be
consistently quantized in the ``traditional'' quantum field theory
way, as gravity cannot be quantized too. Perhaps this is another
hint that makes us speculate about the possible interconnection of
gravity and the neutrino field.

Another quite astonishing feature of tachyonic fermions is that,
if such a particle loses energy in a medium, it will undergo an
acceleration \cite{bilaniuk}. If we assume that the neutrino is a
tachyon, there is no direct way for this particle to loose energy,
due to the fact that it has no electric charge. Nevertheless, it
is known that a neutrino travelling through a medium can have
leading order matter induced magnetic and electric dipole moments
\cite{pal}. The last can be non-zero even for Majorana neutrinos.
It is not likely to directly measure such effects, however if the
neutrino is proven to be tachyon, then these interactions could
make the neutrino loose energy in various dense medium. This could
have observable effects coming from astrophysical sources and
astrophysical events. Since the energy loss of a tachyon causes
it's acceleration, then neutrinos loosing energy in a Gamma Ray
Burst generating background, would eventually accelerate and would
reach earth faster than a photon. This result is similar to the
quantum gravity one, we discussed earlier.

\subsubsection*{Relativistic Quantum Mechanics, Localization and A-causality and Superluminal Propagation}

In this section we shall discuss a curious and rather not so very
well known problem that appears in the relativistic Dirac
equation. We base our analysis in \cite{thaller}. It is known that
any solutions of the Dirac equation cannot describe propagation
faster than light. However in the Foldy-Wouthuysen representation
(a unitary transformation of the Dirac and other operators, that
transforms the Dirac equation into two Klein-Gordon equations),
superluminal propagation is possible, in terms of a superluminal
spreading of the wave functions. This problem is very closely
related to position operators commuting with the positive energy
and in addition to the way we define localization of a particle.
If we assume a quantum mechanical-consistent definition of the
localization of a particle in a subspace $B$ of $R^3$, then there
is a non-zero probability that this particle can be found outside
this space $B$, at a space $B_0$. This instantaneous spreading of
the wave functions corresponds to superluminal propagation. The
whole argument can be summarized in terms of the following
theorem, which states that:

Let $D$ be a Hilbert space. For every set $B\subset R^3$ let
$F(B)$ be a bounded self adjoint operator in $D$, such that for
every function $\psi$ belonging in $D$ with $\langle \psi|
\psi\rangle =1$, the probability measure $\langle
\psi|F(B)|\psi\rangle $ satisfies  the following properties:

\begin{itemize}

\item $\langle \psi|F(B)|\psi\rangle=1$ and $\langle
\phi|F(B)|\phi\rangle=0$ implies $\langle\phi|\psi\rangle=1$

\item There is a self adjoint operator $p$ (the generator of space
translations) such that for all $a$ $\epsilon$ $R^3$,

$\langle e^{-ipa}\psi|F(B+a)|e^{-ipa}\psi\rangle=\langle
\psi|F(B)|\psi\rangle$
\end{itemize}

Furthermore define a non-constant positive operator $H(p)$. For
all non empty sets $B_i\subset R^3$ and for all $\varepsilon >0$,
there exists a time $t$ $\epsilon$ (0,$\varepsilon$), for which

\begin{equation}\label{theoremthatl}
\langle e^{-iHt}\psi|F(B_i)|e^{-iHt}\psi\rangle \neq 0
\end{equation}

Therefore if $F(B_i)$ represents a position operator, there is a
small probability that the particle can be outside $B_i$, to
another subset of $R^3$ for arbitrarily small time. It is expected
that in physically reasonable situations, this instantaneous
spreading of the wave functions must be very small. For a much
more detailed analysis we refer the reader to reference
\cite{thaller} (page 30).

\subsubsection*{An $U(2,2)$-invariant Modification of the Dirac
Equation}

The standard Dirac Lagrangian in curved spacetime is built as a
gauge theory on a manifold $M$, with $SL(2,C)$ structural group.
The vielbeins $e_{\mu}^A$ are the sections of the corresponding
principal bundle, the spin connections $\omega^{r}_{s\mu}$ are
corrections of the principal bundle, and the fields $\psi_r$ are
the sections of the associate bundle with fibre $C^4$. A more
general spinor equation in curved space can be built if we admit
the structural group to be the total pseudo-unitary group
$U(2,2)$, to which $SL(2,C)$ is an injected subgroup
\cite{polish}. The group $U(2,2)$, is used in twistor geometry and
conformal field theory. We shall consider a very simple form of
the modified $U(2,2)$-Dirac Lagrangian (see \cite{polish} for
details), by taking $e^{\mu}_A=\delta^{\mu}_A$, the internal
metric to be locally Euclidean, that is
$g_{\mu,\nu}=\delta_{\mu,\nu}$. The Lagrangian reads,
\begin{equation}\label{modlag1}
\mathcal{L}=V\frac{i}{2}\big{(}\bar{\psi}\gamma^{\mu}\partial_{\mu}\psi-\partial_{\mu}\bar{\psi}\gamma^{\mu}\psi\big{)}-W\bar{\psi}\psi+U\partial^{\mu}\bar{\psi}\partial_{\mu}\psi
\end{equation}
The corresponding equation of motion is,
\begin{equation}\label{modlag}
Vi\gamma^{\mu}\partial_{\mu}\psi-W\psi-U\partial^{\mu}\partial_{\mu}\psi=0
\end{equation}
with $V,W,U$, real constants (some of these have mass dimensions).
The above equation does not correspond to any relativistic
equation of an elementary particle, since no representation of the
Poincare group results in equation (\ref{modlag}). In addition,
the quantization of the above equation is not usual in the
traditional Quantum Field Theory-way because tachyonic solutions
appear, and obviously due to the second derivatives in
(\ref{modlag}). Nevertheless we think that (\ref{modlag})
describes a particle dynamical system not respecting Poincare
invariance in 4-dimensions. The reason for considering such an
physics-unattractive equation, is due to the outcomes, in
reference to tachyonic solutions. Indeed, if the solutions of
(\ref{modlag}) are of the form, $\psi=\phi e^{-ip_{\mu}x^{\mu}}$,
then equation (\ref{modlag}), becomes:
\begin{equation}\label{eqne}
\gamma^{\mu}p_{\mu}\phi=m\phi
\end{equation}
with two solutions for the mass, namely:
\begin{equation}\label{masssol}
m_{1,2}=\Big{(}\frac{1}{2U^2}\big{(}2UW+V^2\pm\sqrt{V^4+4UWV^2}\big{)}\Big{)}^{1/2}
\end{equation}
Therefore, the solutions to (\ref{modlag}), correspond to a
superposition of two Dirac waves with masses $m_1,m_2$. There are
values in the parameter space for which the masses can be purely
imaginary. This means that we may have tachyonic solutions for the
wave particle $\psi$. In addition there is a mass gap between the
two masses. This means that if we choose appropriately the
parameters, we can obtain a large mass gap between the masses
$m_1$ and $m_2$. If we consider that the field $\psi$ could
describe the neutrino, both the results we described above are
quite interesting. Especially the second result, since it is
reminiscent to the phenomenological outcomes of the seesaw
mechanism \cite{seesaw}. Indeed if the two wave solutions with
masses $m_1$ and $m_2$, describe the left handed and right handed
neutrino, we can achieve the mass splitting to be huge, so that
the right handed neutrino's cannot be observed. We must notice
that the aforementioned mass splitting occurs even for real masses
and not only for imaginary masses. This scenario is quite
attractive, in regards to tachyonic solutions, since tachyons
always come in left and right handed components. Hence the above
mechanism could provide a way to split energetically the two
masses, keeping one low and the other one decoupled due to it's
high mass.

\subsubsection*{Conclusions}

In the present paper we studied a toy model of fermions who's main
feature is the dynamical generation of a tachyonic mass for the
fermions involved. No attempt was made to produce phenomenology or
imply that the neutrino indeed is a tachyon. The only assumption
we made is that the toy fermion has a Lagrangian of the form
(\ref{tachyonlagra}). The phenomenological implications of such a
Lagrangian at the Standard Model level are identical to those of a
massless Dirac fermion. It is quite a mystery why it yields
tachyonic mass generation when non-perturbative effects are taken
into account.

Furthermore we discussed various implications of a tachyonic
neutrino in physical processes. We can sum up here the most
important ones:

\begin{itemize}

\item The energy loss of a tachyon results in the acceleration of
the particle. Consequently, neutrinos loosing energy in a Gamma
Ray Burst generating background could reach earth faster than a
photon.

\item In hot-dense backgrounds, the plasmon decay and the
effective neutrino-photon coupling can be modified and therefore
have direct impact on the neutrino emission from stars

\item There are no tachyonic neutrinos which are only left handed.

\end{itemize}

In addition we provided a toy $U(2,2)$ invariant modification of
the Dirac equation (originally presented in \cite{polish} but in a
different context) which can admit tachyonic solutions and can
provide a natural seesaw-like mechanism. However this modification
we provided cannot describe a common sense-acceptable relativistic
particle. It is just a geometrodynamical toy model with
interesting features.

It is tempting to discuss problems such as the chiral symmetry
restoration at finite temperature. Nevertheless it is extremely
difficult to address these problems without knowing how to handle
the resulting extrema. The finite temperature effective potential
of the modified Gross-Neveu model we used is:
\begin{align}\label{newtemp1}
&V_{eff}=-\frac{2\sqrt{\pi }}{(2\pi )^{d}a}(2\pi )^{\frac{d-1}{2}
}m^{d+1}\Gamma (-\nu -\frac{1}{2}+1)(a^{2})^{\frac{1}{2}-\nu
}\\&\notag \times [\sum_{k=1}^{\infty
}\sum_{l=0}^{\lambda }\frac{((2\pi )^{2})^{\nu -\frac{1}{2}-l}(\nu -\frac{1}{2})!}{%
(\nu -\frac{1}{2}-l)!l!}(a^{2})^{l}(k^{2})^{\nu
-\frac{1}{2}-l}]\\&\notag +\frac{1}{2}\frac{\sqrt{\pi }}{(2\pi
)^{d}a_{1}}(2\pi )^{\frac{d-1}{2} }{(m^{3d+3})}^{d+1}\Gamma (-\nu
-\frac{1}{2}+1)-\ \ \frac{1}{4}\frac{1}{(2\pi )^{d}}(2\pi
)^{\frac{d-1}{2}}{(m^{3d+3})}^{d+1}\Gamma (-\nu )\\&\notag
+\frac{1}{2}\frac{\sqrt{\pi }}{(2\pi )^{d}a_{1}}(2\pi
)^{\frac{d-1}{2} }{(m^{3d+3})}^{d+1}\Gamma (-\nu
-\frac{1}{2}+1)(a_{1}^{2})^{\frac{1}{2}-\nu }\\& \times
[\sum_{k=1}^{
\infty }\sum_{l=0}^{\lambda }\frac{((2\pi )^{2})^{\nu -\frac{1}{2}-l}(\nu -\frac{1}{%
2})!}{(\nu -\frac{1}{2}-l)!l!}(a_{1}^{2})^{l}(k^{2})^{\nu
-\frac{1}{2}-l}] ,\end{align} with $\alpha =\frac{m^{-2}}{T}$ and
$\alpha _{1}=\frac{m^{-2}}{2T}$. In addition,
\begin{equation}\label{85}
\nu =\frac{d+1}{2} .\end{equation} and also, $\lambda$ is a
positive integer. In the above, $m$ represents the dynamical
generated minimum of the effective potential $\sigma$. This is to
be found by minimizing the effective potential. Using the zeta
regularization technique
\cite{bordagreview,elizaldenew1,odi,oikonomou} the effective
potential can be cast as,
\begin{align}
&V_{eff} =-\frac{\sqrt{\pi }}{(2\pi )^{d}a}(2\pi )^{\frac{d-1}{2}%
}{(m^{3d+3})}^{d+1}\Gamma (-\nu -\frac{1}{2}+1)+\ \
\frac{1}{2}\frac{1}{(2\pi )^{d}}(2\pi
)^{\frac{d-1}{2}}{(m^{3d+3})}^{d+1}\Gamma (-\nu )\nonumber
\\&\notag -\frac{2\sqrt{\pi }}{(2\pi )^{d}a}(2\pi
)^{\frac{d-1}{2} }{(m^{3d+3})}^{d+1}\Gamma (-\nu
-\frac{1}{2}+1)(a^{2})^{\frac{1}{2}-\nu }\\&\notag \times
[\sum_{l=0}^{\lambda }
\frac{((2\pi )^{2})^{\nu -\frac{1}{2}-l}(\nu -\frac{1}{2})!}{(\nu -\frac{1}{2}-l)!l!}%
(a^{2})^{l}\zeta (-2\nu  +1+2l)]  \\&\notag
+\frac{1}{2}\frac{\sqrt{\pi }}{(2\pi )^{d}a_{1}}(2\pi
)^{\frac{d-1}{2} }{(m^{3d+3})}^{d+1}\Gamma (-\nu -\frac{1}{2}+1)-\
\ \frac{1}{4}\frac{1}{(2\pi )^{d}}(2\pi
)^{\frac{d-1}{2}}{(m^{3d+3})}^{d+1}\Gamma (-\nu ) \\&\notag
+\frac{1}{2}\frac{\sqrt{\pi }}{(2\pi )^{d}a_{1}}(2\pi
)^{\frac{d-1}{2} }{(m^{3d+3})}^{d+1}\Gamma (-\nu
-\frac{1}{2}+1)(a_{1}^{2})^{\frac{1}{2}-\nu }\\&\notag \times
[\sum_{l=0}^{
\lambda }\frac{((2\pi )^{2})^{\nu -\frac{1}{2}-l}(\nu -\frac{1}{2})!}{(\nu -\frac{1}{2}%
-l)!l!}(a_{1}^{2})^{l}\zeta (-2\nu  +1+2l)] .\label{newtemp2}
\end{align}

\noindent We kept the above expression without simplifying in
order to have a clear picture of the terms appearing. In the case
$d=1$  poles appear in the expression above, which are cancelled
using the zeta regularization technique. This can easily be seen
Taylor expanding around $d=1+\epsilon $ for $\epsilon \rightarrow
0$.

It is obvious that any question in reference to chiral symmetry
restoration, corresponds to finding the minimum of the effective
potential above, in terms of $\sigma=m$. The locus of local
extrema of the thermal effective potential is very complex to
handle. Moreover we cannot be sure of what an imaginary solution
for the thermal mass could indicate, therefore in order to  avoid
un-physical speculations we postpone such an investigation. It
would be interesting though to study the finite temperature case
with the appearance of a complex (or not) chemical potential. We
hope to address such problems in the future.

\end{document}